\documentclass[apj]{emulateapj}
\slugcomment{{\sc Accepted to AJ:} November 24, 2009}
\def\mr{\mathrm}

\def\kms{\mr{~km\ s}^{-1} }
\def\hi{H{\sc i}~}

\def\msol{\mr{M}_{\odot}}
\def\sfr{\ \msol\ \mr{yr}^{-1}}

\begin{document}
\title{Evidence for a Strong End-On Bar in the Ringed $\sigma$-Drop Galaxy NGC 6503}
\author{E. Freeland\footnotemark[1], L. Chomiuk, R. Keenan, \& T. Nelson\footnotemark[2]}
\footnotetext[1]{Now at the George P. and Cynthia W. Mitchell Institute for Fundamental Physics and Astronomy, Dept. of Physics and Astronomy, Texas A\&M University}
\footnotetext[2]{Now at the X-ray Astrophysics Laboratory, NASA/GSFC and Dept. of Physics, University of Maryland-Baltimore County}
\affil{Department of Astronomy, University of Wisconsin, Madison, WI 53706}

\begin{abstract}  We use WIYN High-resolution Infrared Camera (WHIRC) $H$ ($1.6\ \mu$m) band imaging, archival \emph{Spitzer}, \emph{HST}, and \emph{GALEX} data, simulations and data from the literature to argue for the presence of a strong end-on bar in nearby spiral galaxy NGC 6503.  The evidence consists of both photometric and kinematic signatures as well as resonant structures present in the galaxy which are most often associated with bars.  These include a central peak followed by a plateau in the surface brightness profile, an extreme decrement in the central stellar velocity dispersion (a $\sigma$-drop), and the presence of an inner ring as well as a circumnuclear disk with spiral structure.  In this framework the previously identified nuclear star-forming ring is instead a young inner ring spanned in diameter by the strong end-on bar. 
\end {abstract}


\section{Introduction}

Near infrared (NIR) observations reveal stellar bars in a majority of disk galaxies \citep{2000AJ....119..536E,2007ApJ...659.1176M}.   Observations at these wavelengths probe older stellar populations (the primary components of bars), and, unlike optical observations, are relatively unaffected by young stars and dust.  Bars play an important role in redirecting gas from the outskirts of galaxies toward small radii by gravitationally torquing the gas, causing it to lose angular momentum and flow radially into the center of the galaxy \citep{1994ApJ...424...84H}.  However, these flows may be stopped by the Inner Lindblad Resonance (ILR), and the gas redistributed into a disk or ring which may form stars \citep{1990Natur.345..679S,1996FCPh...17...95B}.  To fuel an AGN or nuclear starburst, it is then necessary for another mechanism to perturb this gas, causing it to further flow into the inner $\sim10$ pc.  Possible mechanisms include gravitational perturbation from a companion galaxy and instabilities in the disk, including small nuclear bars \citep{1989Natur.338...45S}.  

Some $20 \pm 5\%$ of local spirals have nuclear star-forming rings \citep{2005A&A...429..141K}, and most are thought to be associated with non-axisymmetric structures like bars \citep{2009arXiv0908.0272C,1990Natur.345..679S,1994mtia.conf..143A}.  There are, however, a few examples of disk galaxies with nuclear star-forming rings and no obvious bar structure \citep{2008ApJS..174..337M}.  These cases are often explained by the perturbations from other non-axisymmetric structures such as spiral density waves, oval distortions, or minor mergers \citep{2004A&A...423..481K,2006AJ....131.1336S}.  Spiral density waves can drive gas into the center of galaxies, but this mechanism is not thought to be efficient considering the lack of stability of the wave in the absence of a bar \citep{1996FCPh...17...95B}.  Oval galaxies are thought to evolve similarly to barred galaxies because, although the elongation is weaker than a bar, a larger fraction of the disk mass is affected by the nonaxisymmetry \citep{2004ARA&A..42..603K}.  \citet{1997MNRAS.286..284A} show that the impact of a small satellite can produce a ring that is indistinguishable from a resonance ring.

\citet{2008ApJS..174..337M} survey 22 nearby spirals with nuclear star-forming rings in H$\alpha$, $B$, and $I$ bands and do not find bars in five of these systems.  From this sample of galaxies with star-forming rings but no detected bars, we have chosen NGC 6503 for NIR imaging, with the new WIYN\footnote{The WIYN Observatory is a joint facility of the University of Wisconsin--Madison, Indiana University, Yale University, and the National Optical Astronomy Observatories.} High-resolution InfraRed Camera (WHIRC) on the WIYN 3.5m on Kitt Peak, in order to further search for a bar associated with the star-forming ring.  The highly inclined disk in NGC 6503 makes this source an interesting target for NIR observations.   Table \ref{tab:sour} lists some general properties of this galaxy and its star-forming ring.

NGC 6503 is a spiral galaxy at a distance of 5.3 Mpc, near the edge of the Local Void \citep{2003A&A...398..479K}.  It appears to be extremely isolated with only one possible faint dwarf companion whose radial velocity is unknown \citep{1998A&AS..127..409K,2003A&A...398..479K} and no \hi detections nearby \citep{2009AJ....137.4718G}.  This galaxy provides a unique laboratory in which to study a nearby system which is likely unperturbed by neighbors.  There is evidence for modest gas inflow onto a central black hole, as \citet{2007MNRAS.382.1552L} classify the bright nucleus as a LINER and \citet{1997ApJS..112..315H} use an uncertain designation of transition object or Seyfert 2. It is a low-luminosity spiral galaxy with a star formation rate of $0.2 \sfr$ \citep{1998ARA&A..36..189K}, and an inclination of $74^\circ$ \citep{1997MNRAS.290..585B}.  

NGC 6503 hosts the most extreme case known of a decrement in the stellar velocity dispersion in the center of a galaxy, also known as a $\sigma$-drop \citep{1989A&A...221..236B,1993A&A...275...16B,1997MNRAS.290..585B,2005ApJ...626..159B,2008A&A...485..695C}.  Within the central $\sim 10 \arcsec$ the velocity dispersion drops approximately in half, from $\sim50 \kms$
to $\sim25 \kms$ \citep{1989A&A...221..236B}.  The $\sigma$-drop phenomenon is seen in $30-50\%$ of disc galaxies \citep{2008A&A...485..695C}, and is often attributed to a bar or a circumnuclear disk.  In the case of a bar, stellar orbits in the center of a bar circularize, decreasing the velocity dispersion \citep{2005ApJ...626..159B}.  In the case of a circumnuclear disk, the decrease in velocity dispersion is caused by dynamically cold gas clouds which form a population of stars whose velocity dispersion, like that of the clouds they formed in, is small \citep{2001A&A...368...52E,2003A&A...409..469W}.  This young stellar population dominates the luminosity from the central region of the galaxy and may overpower the light from a physically coincident older population which would not show a $\sigma$-drop.  These two hypotheses are not mutually exclusive, indeed, as discussed earlier, bars are often invoked as a mechanism to transfer gas into the centers of galaxies.  \citet{2006MNRAS.369..853W} find that the timescale to form a $\sigma$-drop in a circumnuclear disk can be as short as $\sim 500$ Myr and the signature will last as long as the gas content of the circumnuclear disk remains high enough that stars continue to form.   
Therefore, it remains a curiosity that NGC 6503 displays \emph{two} signatures which are commonly associated with bars, yet no bar has been detected in this galaxy to date. Here, we use new NIR data, multi-wavelength archival data, and kinematics from the literature to revisit this system and search for evidence of a bar which might simultaneously explain its star-forming ring and strong $\sigma$-drop. 

\begin{deluxetable*}{lccc}
\tablecaption{Basic Data for NGC 6503\label{tab:sour}}
\tabletypesize{\scriptsize}
\tablehead{Parameter & Value & Reference}
\startdata
RA (J2000) & 17h49m26.5s & NED \\
Dec (J2000) & +70d08m40s & NED \\
Morph Type & SA(s)cd;  \ion{H}{2}/LINER & NED \\
Distance & 5.3 Mpc & Karachentsev et al.~2003 \\
Semi-Major Axis (R$_{25}$) & 3.5 arcmin & de Vaucouleurs et al.~1991 \\
Inclination & 74 deg & Bottema \& Gerritsen 1997\\
R$_{ring}$\tablenotemark{a} (angular) & $38.5^{\prime\prime} \times13.4^{\prime\prime}$ & Mazzuca et al.~2008 \\
R$_{ring}$\tablenotemark{b} (physical) & 1.0 kpc $\times$ 0.3 kpc  
\enddata
\tablenotetext{a}{Angular ring dimensions are unprojected.}
\tablenotetext{b}{Physical ring dimensions assume the angular ring dimensions and distance listed in this table.}
\end{deluxetable*}

\section{Observations}

\begin{figure}
\plotone{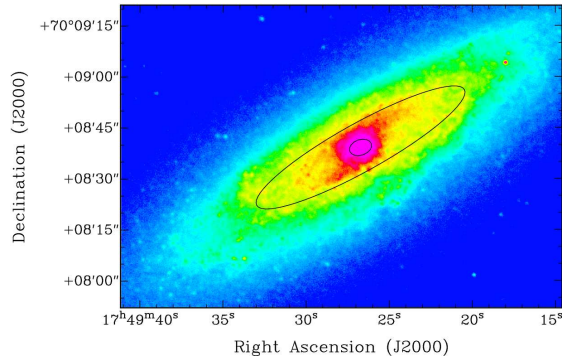}
\caption{$H$-band image of NGC 6503.  The small black ellipse indicates the extent of the circumnuclear disk with a semi-major axis of $3.5 \arcsec$.  The large black  ellipse indicates the inside edge of the star-forming ring with a semi-major axis of $36 \arcsec$.  The strong end-on bar appears round in projection and prominent NIR spiral arms are seen coming off the ends.  }\label{fig:6503}
\end{figure}

We observed NGC 6503 with the WHIRC camera on WIYN during a single night in September 2008.  The WHIRC instrument is a $2048 \times 2048$ HgCdTe infrared array with a field of view of $3\arcmin \times 3\arcmin$ which is capable of sub-arcsecond resolution for broad and narrow band imaging between $0.8$ and $2.5$ $\mu$m \citep{2008SPIE.7014E..96M}. The pixel scale is $0.098 \arcsec~\mathrm{pix}^{-1}$.  The detector is mounted on the WIYN Tip-Tilt Module which enables spatial resolution as good as $0.3\arcsec$ FWHM through active image correction, although this capability was turned off during our observations.  We imaged the galaxy in the $H$ band ($1.6\ \mu$m) which provided the best compromise between sensitivity to the old stellar component and minimization of the sky background.  In order to facilitate sky subtraction, we obtained a series of images both on and off the source with integration times ranging from 50 to 200 s. The data were reduced in IDL following a procedure recommended by the WHIRC science team (Dick Joyce, private communication 2008).  All image frames were bias-corrected and flat-fielded.  A linearity correction was applied using the coefficients listed in the WHIRC manual (\verb|www.noao.edu/kpno/manuals/whirc/WHIRC.htm|).  A master background frame was produced by median combining the off-source images and then subtracted from each image of the galaxy.  Finally, the object images were stacked to produce the $H$-band image shown in Figure \ref{fig:6503}.  The stacked image of NGC 6503 has a final FWHM seeing of $0.9\arcsec$ and a combined exposure time of 1500 seconds.

In addition to the WHIRC NIR data we retrieved coadded near-ultraviolet (NUV) observations of this galaxy from the \emph{GALEX} archive (PI Hoopes, Proposal ID 97) which have a combined exposure time of 4300 sec.  Far-ultraviolet (FUV) survey observations were available with a combined exposure time of 1600 sec.  We also retrieved archival \emph{Hubble Space Telescope (HST)} observations in the F814W and F658N filters with exposures time of 700 seconds and 1700 seconds, respectively (PI Ford, Proposal ID 9293) and public \emph{Spitzer Space Telescope} IRAC images at 3.6 and 4.5 $\mu$m (PI Leitherer, Proposal ID 3674).  \citet{2009AJ....137.4718G} provided us with their total intensity \hi image from the Very Large Array.\footnote{The Very Large Array is operated by the the National Radio Astronomy Observatory, which is a facility of the National Science Foundation operated under cooperative agreement by Associated Universities, Inc.}  This C-configuration radio data has a resolution of $14.4 \arcsec \times 13.5 \arcsec$ with $500$ minutes on source.

\section{The Strong End-on Bar}

\citet{2005ApJ...626..159B} perform self-consistent three-dimensional N-body simulations of disks with bars of varying strength and orientation in order to identify the signatures of bars in highly inclined galaxies.  These simulations contain only a luminous disk and dark matter halo; no bulge component is included.  Bar diagnostics arise from simulated long-slit spectra along the major axis of the disk and include position-velocity diagrams, surface brightness profiles, velocity profiles, velocity dispersion profiles, and the Gauss-Hermite terms $h_3$ and $h_4$.  Here, we compare observations of NGC 6503 to these theoretically predicted features using our WHIRC data as well as the following data from the literature.  When comparing surface brightness profiles we use $B$ and $R$-band data published in \citet{1989A&A...221..236B} and the 3.6 $\mu$m and 4.5 $\mu$m data from the \emph{Spitzer} archive.  Rotation curves for NGC 6503 have been published using H$\beta$ data in \citet{1989A&A...221..236B} and H$\alpha$ data in \citet{1982ApJS...49..515D}.  The stellar velocity dispersion profile was first presented in \citet{1989A&A...221..236B}. 

First, we apply the commonly used method of ellipse fitting to search for a bar in our WHIRC data; for more details see \citet{2007ApJ...657..790M}.  This method looks for specific signatures in the ellipticity ($\epsilon = 1 - b/a$, where $b$ and $a$ are the semi-minor and semi-major axes, respectively) and position angle (PA) of the galaxy isophotes as a function of increasing semi-major axis in the galaxy disk.  The ellipticity should increase monotonically along the length of the bar while the PA stays relatively constant.  At the end of the bar there is a drop in the ellipticity and a correspondingly abrupt change in the PA as the isophotes round out into a disk.  These signatures are due to the shape and orientation of stellar orbits in the bar \citep{1992MNRAS.259..328A}.  The existence of spiral arms will cause the PA to change slowly instead of remaining constant.  

The IRAF task \verb|ellipse| was used to fit elliptical isophotes to our $H$-band image.  We used a linear step size that was comparable to the resolution of the image.  We performed fits with the center fixed and unfixed and found similar results in both cases.  We created models from our isophote fitting using the IRAF task \verb|bmodel| and subtracted them from our original image to examine the residuals.  The residual image showed no outstanding structure.  

\begin{figure}
\plotone{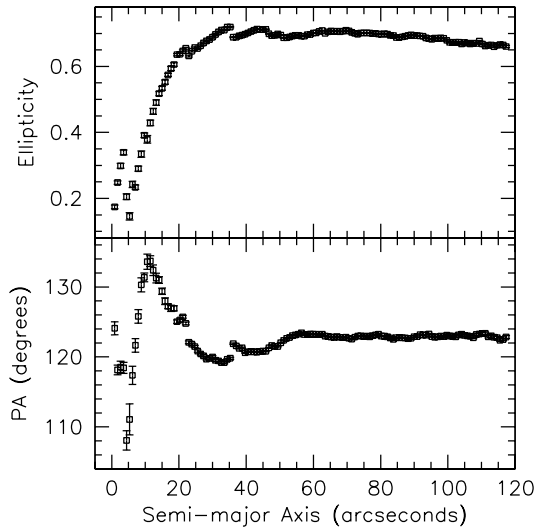}
\caption{Results of the ellipse fitting to the $H$ band image of NGC 6503. Ellipticity as a function of semi-major axis in arcseconds is shown in the top panel, and position angle is shown in the lower panel.  Points are spaced by a single resolution element, in this case $0.9 \arcsec$.  Error bars reflect the errors in the harmonic fit after the first and second harmonics have been removed \citep[see][]{1996ASPC..101..139B}.  }\label{fig:6503epa}
\end{figure}

\begin{figure}
\plotone{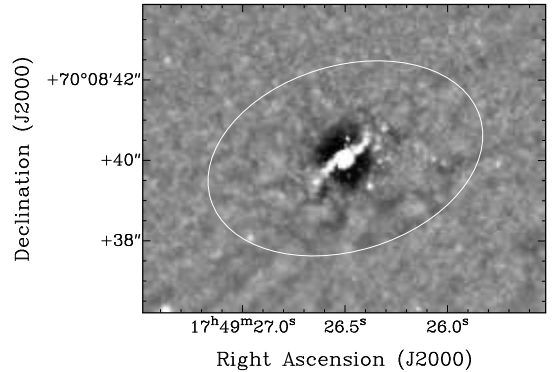}
\caption{Unsharp-masked \emph{HST} F814W image (smoothed at a scale of $0.3 \arcsec$) showing the nuclear spiral structure.  The white ellipse marks the extent of the circumnuclear disk according to the ellipse fitting.}\label{fig:nucsp}
\end{figure}

\begin{figure*}[htp]
\plotone{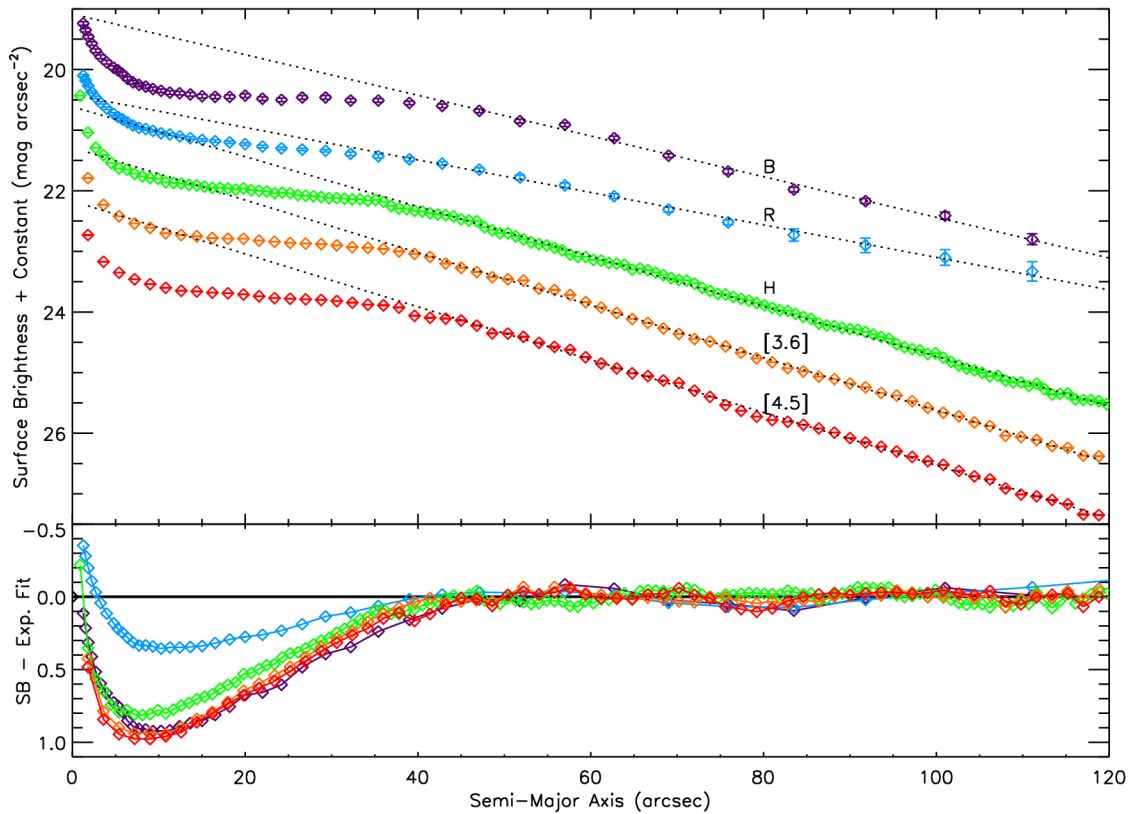}
\caption{Top panel: surface brightness profiles in arbitrary units of mag arcsec$^{-2}$ measured at five different bands: $B$ and $R$ from \citet{1989A&A...221..236B}, WHIRC $H$, and \emph{Spitzer} IRAC bands at 3.6 $\mu$m and 4.5 $\mu$m. At each band, an exponential profile is fit to semi-major axes between 45$^{\prime\prime}$ and 120$^{\prime\prime}$; these fits are shown as black dotted lines. Bottom panel: the residuals in the surface brightness profiles after subtraction of the exponential fits. Each band is plotted and color-coded as in the top panel.}\label{isocompare}
\end{figure*}

The ellipse fits reveal interesting structure in the ellipticity and PA profiles at small radii, shown in Figure \ref{fig:6503epa}.  The ellipticity rises from the center of the galaxy monotonically while the PA remains relatively constant.  Both the ellipticity and PA change abruptly at a radius of $3.5 \arcsec$.  These are the standard signatures of a bar as outlined above.  However, to unambiguously be considered a bar, the ellipticity needs to reach a global maximum within this structure; we see that the central structure in NGC 6503 only reaches a maximum ellipticity of $\epsilon = 0.35$, as compared with the outer disk which displays $\epsilon = 0.7$. In addition, the PA of the central structure (119$^{\circ}$) is closely aligned to that of the disk (122$^{\circ}$). These characteristics match the criteria for an inner disk laid out by \cite{Erwin_Sparke02}, and we therefore interpret this structure as a circumnuclear disk with a projected angular radius of $3.5\pm0.5 \arcsec$ or a physical length of $89 \pm13$ pc.  Unsharp masking of the archival \emph{HST} F814W image of NGC 6503 reveals a nuclear spiral with a size such that it fits inside the circumnuclear disk (see Figure \ref{fig:nucsp}).  The circumnuclear disk also appears to have a scale similar to that of the sharp central peak in the surface brightness profile, which is also derived from ellipse fitting and shown in Figure \ref{isocompare}. 
 
Moving out in radius past the circumnuclear disk, the PA profile continues to rise till it reaches a maximum value at a radius of $\sim 11 \arcsec$.  This gradual change in position angle is a result of the prominent NIR spiral arms that are visible in the $H$-band image (Figure \ref{fig:6503}).    

Our $H$-band surface brightness profile is plotted in Figure \ref{isocompare} and compared with four other wavelength bands: $B$- and $R$-band data published in \citet{1989A&A...221..236B} and the 3.6 $\mu$m and 4.5 $\mu$m data from the \emph{Spitzer} archive. The profiles in all five bands are Freeman Type II \citep{1970ApJ...160..811F} ---namely, NGC 6503 is well-fit by an exponential profile at large radii ($> 40^{\prime\prime}$), the surface brightness profile almost completely levels out at intermediate radii ($10^{\prime\prime} < r < 40^{\prime\prime}$), and then it increases sharply at small radii ($< 10^{\prime\prime}$).  \citet{2004AJ....127.2085A} find surface brightness profiles of this type in barred galaxies at four times the rate that they are seen in non-barred systems.  \citet{1989A&A...221..236B} claims that the $B$-band surface brightness profile shows a more dramatic plateau than the $R$-band profile, and that the plateau is characterized by redder colors than the rest of the galaxy. They therefore attribute the surface brightness plateau to dust extinction which is concentrated in an annulus at intermediate radii. However, in Figure \ref{isocompare}, we see that the plateau is still clearly visible in the IR bands, which should be significantly less affected by dust. The bottom panel of Figure \ref{isocompare} shows the residuals after an exponential fit has been subtracted from the surface brightness profiles; the plateau regions lie below the $y=0$ line at radii $\lesssim 40^{\prime\prime}$. We see that these residuals have equivalent amplitudes in the $B$, $H$, [3.6], and [4.5] bands (the \citealt{1989A&A...221..236B} $R$-band data are likely affected by additional contamination from H-alpha emission which is quite strong in the inner $40 \arcsec$ of NGC 6503), therefore implying that the plateau remains strong in the IR and excluding dust extinction as its cause. 

An alternative explanation for this surface brightness plateau is a bar. In the simulations of \citet{2005ApJ...626..159B}, galaxies with a wide range of bar strengths, orientations, and inclinations display surface brightness profiles with quasi-exponential central peaks and a plateau at moderate radii.  We note that the plateau extends out to a semi-major axis of $40^{\prime\prime}$, which is strikingly similar to the semi-major axis of the star forming ring at $37.5^{\prime\prime}$ \citep{2008ApJS..174..337M}.  These spatial coincidences are consistent with the simulations of bars by \citet{2005ApJ...626..159B}, which predict that the lengths of features in the rotation curve, velocity dispersion profile, and surface brightness profile are correlated to each other and often to physical features in the galaxy.  

As mentioned in the introduction, NGC 6503 hosts a strong $\sigma$-drop. A strong end-on bar is the only configuration capable of producing a $\sigma$-drop in the simulations of \citet{2005ApJ...626..159B}.  In general, for strong bars a variety of orbital shapes are present along the major axis, and this leads to an increase in the observed velocity dispersion.  However, the orbits near the center of the bar circularize if the potential is deep enough, decreasing the velocity dispersion.  As mentioned in the introduction, a circumnuclear disk containing dynamically cold gas and young stars can also produce a $\sigma$-drop.  Tying these two causes together, the bar may be the cause of the cold disk mentioned above as it funnels gas into the center of the galaxy increasing the strength of the potential.  This combination can further increase the depth of the $\sigma$-drop.  In the case of NGC 6503 we anticipate that circularized stellar orbits in the bar as well as the circumnuclear disk with the nuclear spiral are creating the extreme $\sigma$-drop.  

\begin{figure}
\plotone{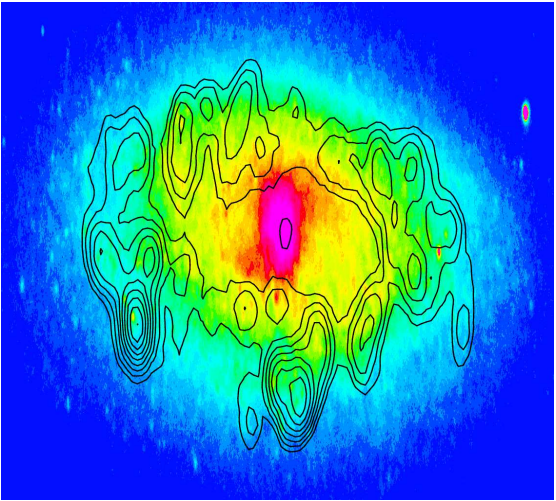}
\caption{Deprojected H-band image of NGC 6503 (color) with deprojected \emph{GALEX} NUV data overlaid (black contours) highlighting the star-forming ring and LINER nucleus (UV point source at the center of the ring).  The NIR spiral arms can be seen extending from the ends of the bar.}\label{fig:deproj}
\end{figure}

Nuclear spirals and star-forming rings are common resonance features.  It is likely that the nuclear spiral and disk occur at the Inner Lindblad Resonance (ILR)  and the star-forming inner ring occurs at the 4:1 ultra-harmonic resonance.  A bar spanning the star-forming inner ring is apparent if we deproject our $H$-band image (shown in Figure \ref{fig:deproj} overlaid with deprojected NUV \emph{GALEX} contours to highlight the ring). The bar has a projected length of $7 \arcsec$;  If we deproject it according to the $74^\circ$ inclination, the true bar length is $25 \arcsec$ or $660$ pc.  The NIR spiral arms can be clearly seen extending from the ends of the bar.   

\section{The Star-Forming Ring} 

\begin{figure*} 
\plotone{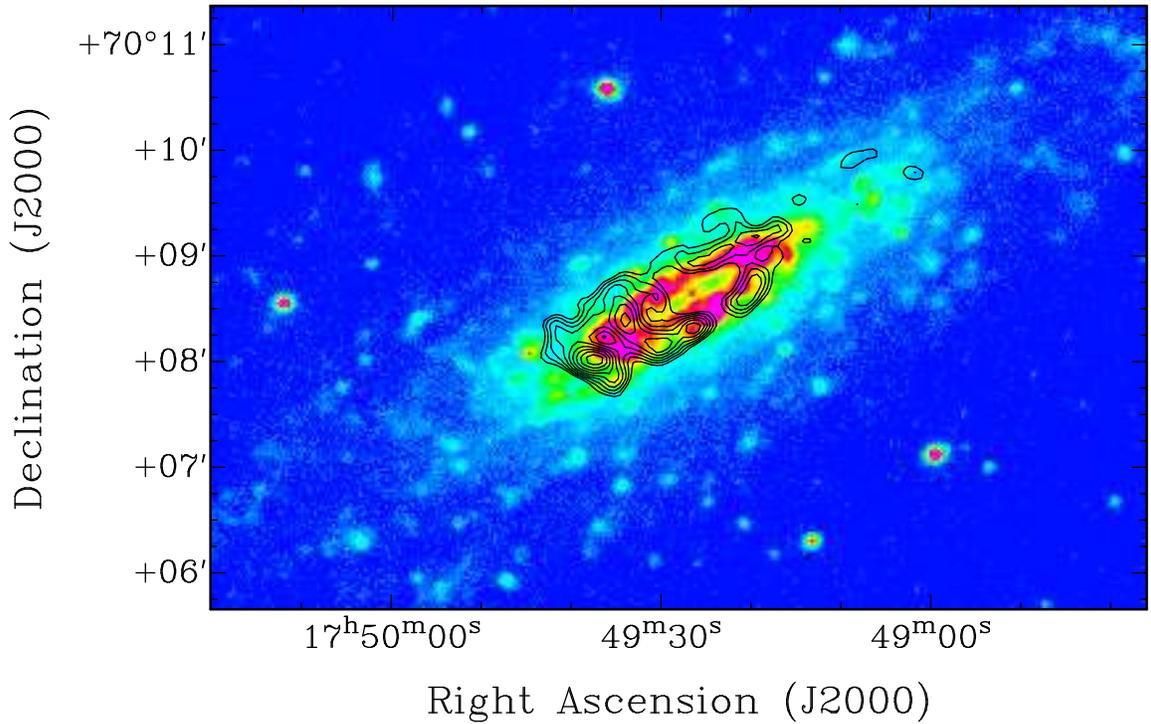} 
\caption{\hi contours (black) overlaid on a \emph{GALEX} NUV image of the center of NGC 6503.  The lowest contour corresponds to gas with a column density of $2 \times 10^{21}\ \mathrm{cm}^{-2}$ (uncorrected for inclination) and contours increase by a factor of 1.05.  We show only the brightest \hi in the galaxy here, the overall \hi content of the galaxy extends to radii of $17-22$ kpc and column densities factors of $100$ times lower \citep[for details of the \hi data see][]{2009AJ....137.4718G}.  The beam size for the \hi data is $14.4 \arcsec \times 13.5 \arcsec$.  At the distance of NGC 6503, $1 \arcmin$ corresponds to $1.5$ kpc.}\label{fig:hi_sf} 
\end{figure*}

The star-forming ring, imaged previously in H$\alpha$ \citep{2006A&A...448..489K}, is more clearly visible in the UV.  Additionally, using the \hi data presented in \citet{2009AJ....137.4718G}, we have identified an \hi ring which is coincident with the star-forming ring; both features are shown in Figure \ref{fig:hi_sf}.  The \hi ring contains $2 \times 10^8\ \msol$ of neutral gas.  An interesting feature of the \hi ring is the lack of gas at the two end-points where the bar intersects the ring.  This is in contrast to inner rings seen in both molecular and neutral gas which do sometimes show enhancements in their gas content at bar ends (e.g. NGC 3627, Maffei 2, NGC 2903, and NGC 6951 in \citealt{2007PASJ...59..117K} and NGC $1433$ in \citealt{1996ApJ...460..665R}).  In NGC 3627 the enhanced molecular gas at the end of the bar appears to be kinematically distinct from the rest of the gas in the ring \citep{2002ApJ...574..126R}.  Hydrodynamical simulations of the gas flow associated with a strong bar show disk gas moving slowly through the regions of the inner ring near the ends of the bar \citep{2008ApJ...684.1048L} as well as piling up and shocking on either side of the bar ends.  The \hi may be missing from the inner ring at the bar ends in NGC 6503 because of recent star-formation which has temporarily depleted or expelled it from those regions.  

\begin{figure*}
\plottwo{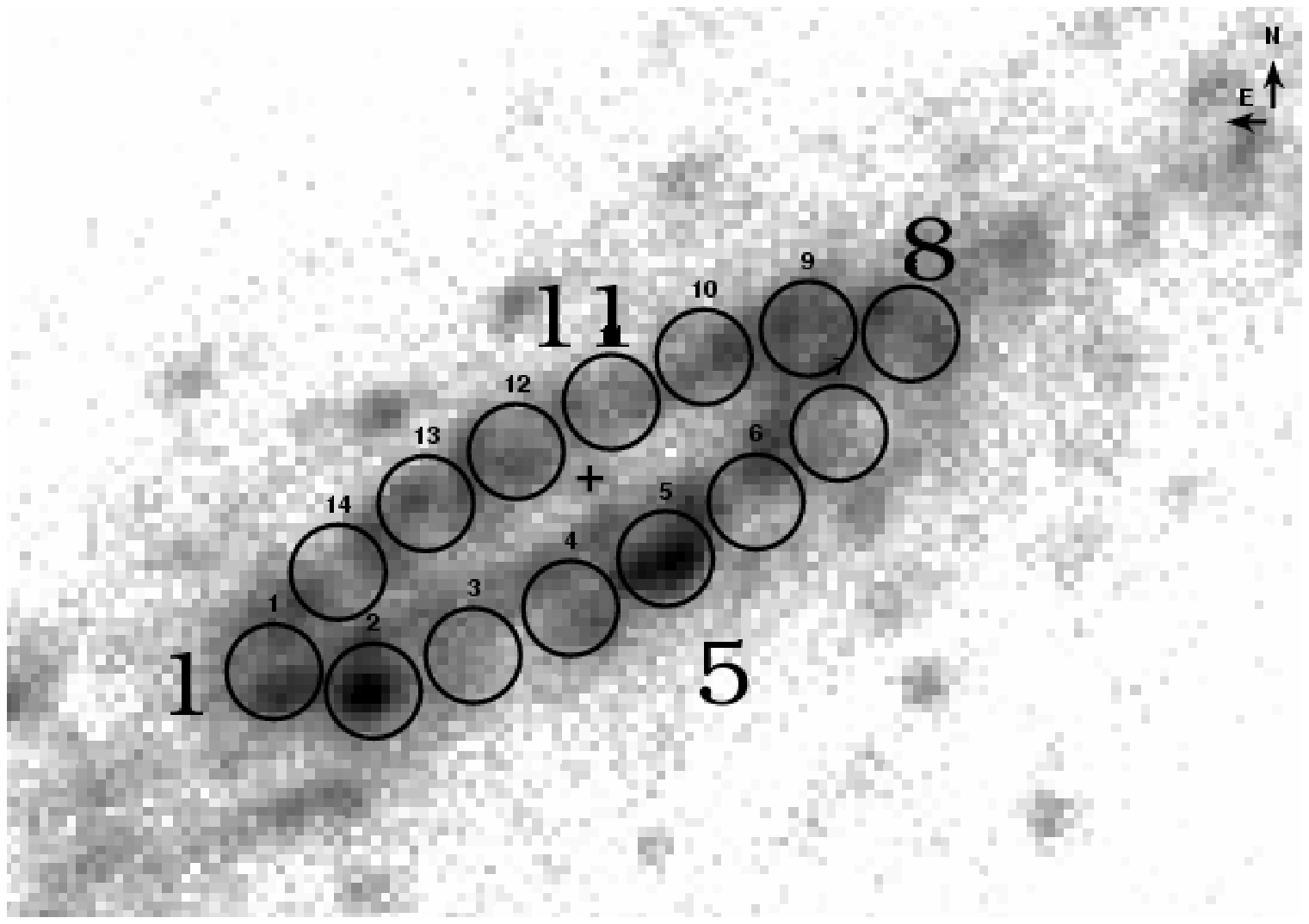}{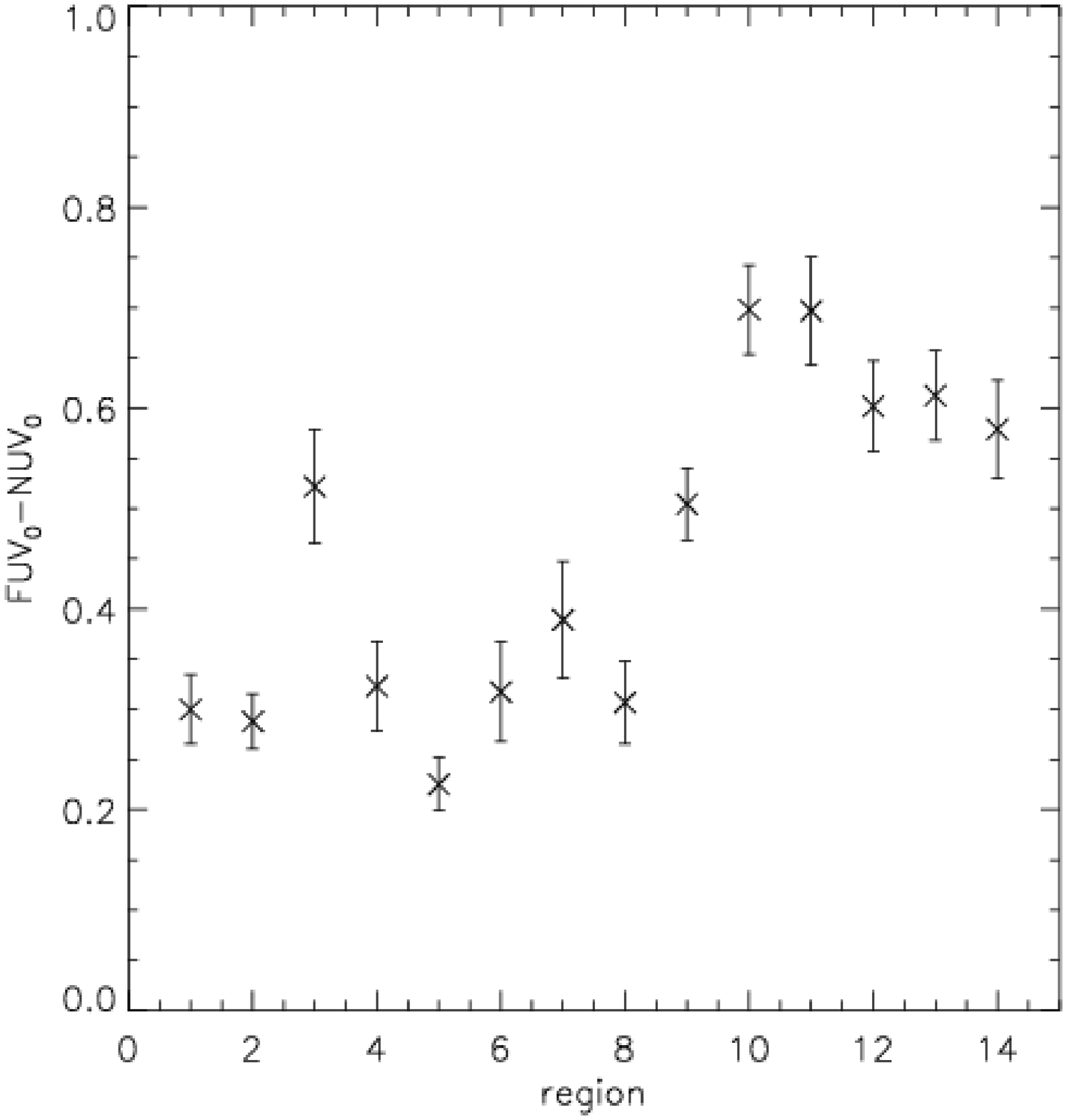}
\caption{Left: Regions selected from the FUV image of the star-forming ring and labelled by number.  Right: FUV-NUV color by region, corrected for reddening due to neutral hydrogen.  The regions on the bottom half of the ring are bluer than those on the top. We interpret these differences seen in FUV-NUV color as being caused by additional extinction (see discussion in \S4).}\label{fig:uvring}
\end{figure*}

We look for gradients in the age of the star-formation around the ring which would identify where gas has most recently entered the ring and formed stars.  For nuclear rings these are called the ``contact points'' and they typically occur in the ring at points perpendicular to the bar major axis \citep{2008ApJS..174..337M}.  There is some evidence for this in \emph{Infrared Space Observatory (ISO)} data at $12\ \mu$m, which shows a ring of hot dust coincident with the star-forming ring and whose brightest emission is seen along the major axis of the galaxy at points nearly perpendicular to the bar major axis \citep{2002AJ....123.3067B}.  Additionally, the \emph{HST} narrow-band H$\alpha$ image shows similar structure with the brightest H{\sc ii} regions seen along the major axis of NGC 6503.  We attempt to quantify any existing age gradient using UV colors.  The \hi data are used to correct the UV data for extinction using the conversion of hydrogen column density to optical extinction from \citet{1995A&A...293..889P} and the relationship between optical reddening and UV reddening from \citet{2005ApJ...619L.119R}. Figure \ref{fig:uvring} shows the FUV$-$NUV color in regions around the ring.  Close examination of dust lanes in the \emph{HST} image indicates that the regions marked $1$--$8$ in the left panel of \ref{fig:uvring} are located on the near side of the galaxy.  These regions appear to have bluer UV colors than those on the far side of the galaxy.  We interpret these differences in FUV$-$NUV color as being caused by additional extinction which is hard to account for entirely in such an inclined system.  Due to the location and size of the bar proposed here, the star-forming ring is much more likely to be an inner ring than a nuclear ring (inner rings are spanned in diameter by the length of the bar whereas nuclear rings are contained within the bar). Thus the physical mechanism that is thought to produce contact points in nuclear rings is not likely to be present for this structure, and if the youngest stars are indeed located perpendicular to the bar major axis this characteristic should be attributed to the dynamics of inner rings. 

The star-forming ring is a young structure and not strongly visible in our WHIRC NIR data which traces older stellar populations.  The UV colors (FUV$-$NUV $< 1$) of regions in the ring indicate that its age is less than $\sim 0.5$ Gyr \citep[see Figure 3 in][]{2007MNRAS.376.1021J}.

\section{Summary}

The star-forming ring in the nearby low-luminosity spiral galaxy NGC 6503 is likely an inner-ring and not a nuclear ring.  We present evidence that the ring is caused by a strong end-on bar which is embedded inside it.  Although the surface brightness profile does not argue exclusively for a strong end-on bar, in the simulations of \citet{2005ApJ...626..159B} a strong end-on bar is the only configuration capable of producing a central $\sigma$-drop.  While it is likely possible to mimic the surface brightness plateau and the central velocity dispersion minimum with axisymmetric structures, we consider the combination and spatial coincidence of these characteristics with resonant features (specifically the inner star-forming ring and circumnuclear disk with spiral structure) to be compelling evidence of a strong end-on bar in this galaxy.   

These results can be tested further with high quality long-slit stellar spectroscopy.  Construction of the rotation curve and velocity dispersion profile with better resolution data, as well as the Gauss-Hermite terms, would allow for futher comparison with the predictions given in \citet{2005ApJ...626..159B}.  Specifically, a strong end-on bar should show a ``double-hump'' signature in its rotation curve such that the rotational velocities rise steeply but then experience a local minimum before rising again to flatten out at large radii.  There may be some evidence for the local minimum in the rotation curves presented in \citet{1982ApJS...49..515D} and \citet{1989A&A...221..236B}, but the data are not adequate to tell conclusively.  Additionally, the velocity dispersion profile for a strong end-on bar should show not only a broad central maximum with a $\sigma$-drop, but also a sharp drop at larger radii and a secondary local maximum.

\vspace{0.5in}

\acknowledgements 

We would like to thank Peter Erwin for his insight as well as Eric Greisen and
Kristine Spekkens for sharing their \hi data with us.  We are extremely
grateful to an anonymous referee, whose thoughtful comments have greatly
improved this paper.  We would like to thank Jere Fluno for his financial
support of this project.  We also appreciate the University of Wisconsin WIYN
Grad Queue and the opportunities it has afforded graduate students over the
years.  We gratefully acknowledge support from NSF grants AST 0709356 and AST
0708002.  This research has made use of the NASA/IPAC extragalactic database
(NED) which is operated by the Jet Propulsion Laboratory, Caltech, under
contract with the National Aeronautics and Space Administration.

\end{document}